\documentclass[3p,times,onecolumn,number]{elsarticle}
\usepackage{amssymb}
\usepackage{amsmath}
\usepackage[dvipdfm]{hyperref}
\usepackage[dvips]{color}
\usepackage{psfig}
\usepackage{subfigure}
\usepackage{float}

\newcommand{\bear}{\begin{eqnarray}}
\newcommand{\eear}{\end{eqnarray}}
\newcommand{\be}{\begin{equation}}
\newcommand{\ee}{\end{equation}}
\newcommand{\beqn}{\begin{eqnarray}}
\newcommand{\eeqn}{\end{eqnarray}}
\newcommand{\beqnn}{\begin{eqnarray*}}
\newcommand{\eeqnn}{\end{eqnarray*}}

\journal{...}

\begin{document}

\begin{frontmatter}
\title{Hole burning in a nanomechanical resonator coupled to a Cooper pair box}

\author[al,CV,BB]{C.~Valverde\corref{cor1}}
\ead{valverde@unip.br}

\author[BB]{A.T.~Avelar},

\author[BB]{B.~Baseia}

\cortext[cor1]{Corresponding author}

\address[al]{Universidade Paulista, Rod. BR 153, km 7, 74845-090 Goi\^ania, GO, Brazil.}

\address[CV]{Universidade Estadual de Goi\'as, Rod. BR 153, 3105, 75132-903 An\'apolis, GO, Brazil.}

\address[BB]{Instituto de F\'{\i}sica, Universidade Federal de Goi\'as, 74001-970 Goi\^ania, GO, Brazil.}

\begin{abstract}
We propose a scheme to create holes in the statistical distribution of excitations of a nanomechanical resonator. It employs a controllable coupling between this system and a Cooper pair box. The success probability and the fidelity are calculated and compared with those obtained in the atom-field system via distinct schemes. As an application we show how to use the hole-burning scheme to prepare (low excited) Fock states.

\end{abstract}

\begin{keyword}
Quantum state engineering \sep
Superconducting circuits  \sep
Nanomechanical Resonator\sep
Cooper Pair Box

\PACS   03.67.Lx, 85.85.+j, 85.25.C, 32. 80. Bx \sep 42.50.Dv
\end{keyword}
\end{frontmatter}

\section{Introduction}

Nanomechanical resonators (NR) have been studied in a diversity of
situations, as for weak force detections \cite{Bocko96}, precision
measurements \cite{Munro02}, quantum information processing \cite{Cleland04}%
, etc. The demonstration of the quantum nature of mechanical and
micromechanical devices is a pursued target; for example, manifestations of
purely nonclassical behavior in a linear resonator should exhibit energy
quantization, the appearance of Fock states, quantum limited
position-momentum uncertainty, superposition and entangled states, etc. NR
can now be fabricated with fundamental vibrational mode frequencies in the
range MHz -- GHz \cite{a1,a2,huang}. Advances in the development of
micromechanical devices also raise the fundamental question of whether such
systems that contain a macroscopic number of atoms will exhibit quantum
behavior. Due to their sizes, quantum behavior in micromechanical systems
will be strongly influenced by interactions with the environment and the
existence of an experimentally accessible quantum regime will depend on the
rate at which decoherence occurs \cite{a3,aaa}. One crucial step in the
study of nanomechanical systems is the engineering and detection of quantum
effects of the mechanical modes. This can be achieved by connecting the
resonators with solid-state electronic devices \cite{a4,a5,a6,a7,a8}, such
as a single-electron transistor. NR has also been used to study quantum
nondemolition measurement \cite{a8,a9,a10,a11}, quantum decoherence \cite%
{a7,a13}, and macroscopic quantum coherence phenomena \cite{a14}. The fast
advance in the tecnique of fabrication in nanotecnology implied great
interest in the study of the NR system in view of its potential modern
applications, \ as a sensor, largely used in various domains, as in biology,
astronomy, quantum computation \cite{a114,a1144}, and more recently in
quantum information \cite{Cleland04,a15,a16,a16a,a17,a19,a20} to implement
the quantum qubit \cite{a16}, multiqubit \cite{a16aa} and to explore cooling
mechanisms \cite{a21,a22,a23,a24,a25,a25a}, transducer techniques \cite%
{a26,a27,a28}, and generation of nonclassical states, as Fock \cite{a29},
Schr\"{o}dinger-\textquotedblleft cat\textquotedblright\ \cite{a7,a30,cv},
squeezed states \cite{a31,a32,a33,a34,a344}, including intermediate and
other superposition states \cite{cv1,cv2}. In particular, NR coupled with
superconducting charge qubits has been used to generate entangled states
\cite{a7,a30,a35,a36}. In a previous paper Zhou and Mizel \cite{a34}
proposed a scheme to create squeezed states in a NR coupled to Cooper pair
box (CPB) qubit; in it the NR-CPB coupling is controllable. Such a control
comes from the change of external parameters and plays an important role in
quantum computation, allowing us to set ON and OFF the interaction between
systems on demand.

Now, the storage of optical data and communications using basic processes
belonging to the domain of the quantum physics have been a subject of
growing interest in recent years \cite{blais}. Concerned with this interest,
we present here a feasible experimental scheme to create holes in the
statistical distribution of excitations of a coherent state previously
prepared in a NR. In this proposal the coupling between the NR and the CPB
can be controlled continuously by tuning two external biasing fluxes. The
motivation is inspired by early investigations on the production of new
materials possessing holes in their fluorescent spectra \cite{a1a} and also
inspired by previous works of ours, in which we have used alternative
systems and schemes to attain this goal \cite{malboi, ard1, ard}. The
desired goal in producing holes with controlled positions in the number
space is their possible application in quantum computation, quantum
cryptography, and quantum communication. As argued in \cite{ard1}, these
states are potential candidates for optical data storage, each hole being
associated with some signal (say YES, $\left\vert 1\right\rangle $, or $%
\left\vert +\right\rangle $) and its absence being associated with an
opposite signal (NO, $\left\vert 0\right\rangle $, or $\left\vert
-\right\rangle $). Generation of such holes has been treated in the contexts
of cavity-QED \cite{ard} and traveling waves \cite{av}.

\section{Model hamiltonian for the CPB-NR system}

\begin{figure}[tbh]
\centering  % figura centralizada
\includegraphics[width=.35\textwidth]{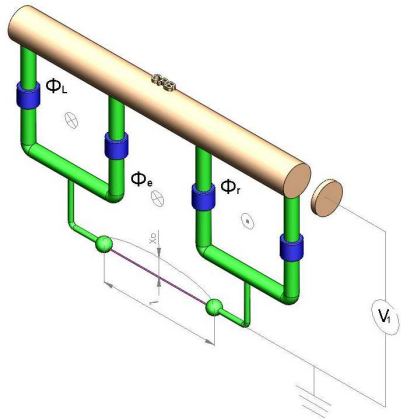}
\caption{\textit{Model for the CPB-NMR coupling.}}
\label{cooper}
\end{figure}

There exist in the literature a large number of devices using the
SQUID-base, where the CPB charge qubit consists of two superconducting
Josephson junctions in a loop. In the present model a CPB is coupled to a NR
as shown in Fig. (\ref{cooper}); the scheme is inspired in the works by
Jie-Qiao Liao et al. \cite{a16a} and Zhou et al. \cite{a34} where we have
substituted each Josephson junction by two of them. This creates a new
configuration including a third loop. A superconducting CPB charge qubit is
adjusted via a voltage $V_{1}$\ at the system input and a capacitance $C_{1}$%
. We want the scheme ataining an efficient tunneling effect for the
Josephson energy. In Fig.(\ref{cooper}) we observe three loops: one great
loop between two small ones. This makes it easier controlling the external
parameters of the system since the control mechanism includes the input
voltage $V_{1}$ plus three external fluxes $\Phi (\ell ),$ $\Phi (r)$ and $%
\Phi _{e}(t)$. In this way one can induce small neighboring loops\emph{.}
The great loop contains the NR and its effective area in the center of the
apparatus changes as the NR oscillates, which creates an external flux $\Phi
_{e}(t)$ that provides the CPB-NR coupling to the system. %
%-----------------------Inicio da figura ------------------------------------------
%-----------------------fim da figura------------------------------------------
%
In this work we will assume the four Josephson junctions being identical,
with the same Josephson energy $E_{J}^{0}$, the same being assumed for the
external fluxes $\Phi (\ell )$ and $\Phi (r)$, i.e., with same magnitude,
but opposite sign: $\Phi (\ell )=-\Phi (r)=\Phi (x)$. In this way, we can
write the Hamiltonian describing the entire system as

\begin{equation}
\hat{H}=\omega \hat{a}^{\dagger }\hat{a}+4E_{c}\left( N_{1}-\frac{1}{2}%
\right) \hat{\sigma}_{z}-4E_{J}^{0}\cos \left( \frac{\pi \Phi _{x}}{\Phi _{0}%
}\right) \cos \left( \frac{\pi \Phi _{e}}{\Phi _{0}}\right) \hat{\sigma}_{x},
\label{a1}
\end{equation}%
where $\hat{a}^{\dagger }(\hat{a})$ is the creation (annihilation) operator
for the excitation in the NR, corresponding with the frequency $\omega $ \
and mass $m$; $E_{J}^{0}$ and $E_{c}$ are respectively the energy of each
Josephson junction and the charge energy of a single electron; $C_{1}$ and $%
C_{J}^{0}$ stand for the input capacitance and the capacitance of each
Josephson tunel, respectively $\Phi _{0}=h/2e$ is the quantum flux and $%
N_{1}=C_{1}V_{1}/2e$ is the charge number in the input with the input
voltage $V_{1}$. We have used the Pauli matrices to describe our system
operators, where the states $\left\vert g\right\rangle $ and $\left\vert
e\right\rangle $ (or 0 and 1) represent the number of extra Cooper pairs in
the superconduting island. We have: $\hat{\sigma}_{z}=\left\vert
g\right\rangle \left\langle g\right\vert -\left\vert e\right\rangle
\left\langle e\right\vert $, $\hat{\sigma}_{x}=\left\vert g\right\rangle
\left\langle e\right\vert -\left\vert e\right\rangle \left\langle
g\right\vert $ and $E_{C}=e^{2}/\left( C_{1}+4C_{J}^{0}\right) .$

The magnectic flux can be written as the sum of two terms,
\begin{equation}
\Phi _{e}=\Phi _{b}+B\ell \hat{x}\text{ },  \label{a4}
\end{equation}%
where the first term $\Phi _{b}$ is the induced flux, corresponding to the
equilibrium position of the NR and the second term describes the
contribution due to the vibration of the NR; $B$ represents the magnectic
field created in the loop. We have assumed the displacement $\hat{x}$
described as $\hat{x}=x_{0}(\hat{a}^{\dagger }+\hat{a})$, where $x_{0}=\sqrt{%
m\omega /2}$ is the amplitude of the oscillation.

Substituting the Eq.(\ref{a4}) in Eq.(\ref{a1}) and controlling the flux $%
\Phi _{b}$ we can adjust $\cos \left( \frac{\pi \Phi _{b}}{\Phi _{0}}\right)
=0$ to obtain
\begin{equation}
\hat{H}=\omega \hat{a}^{\dagger }\hat{a}+4E_{c}\left( N_{1}-\frac{1}{2}%
\right) \hat{\sigma}_{z}-4E_{J}^{0}\cos \left( \frac{\pi \Phi _{x}}{\Phi _{0}%
}\right) \sin \left( \frac{\pi B\ell \hat{x}}{\Phi _{0}}\right) \hat{\sigma}%
_{x},  \label{a8}
\end{equation}%
and making the approximation $\pi B\ell x/\Phi _{0}<<1$ we find
\begin{equation}
\hat{H}=\omega \hat{a}^{\dagger }\hat{a}+\frac{1}{2}\omega _{0}\hat{\sigma}%
_{z}+\lambda _{0}(\hat{a}^{\dagger }+\hat{a})\hat{\sigma}_{x},  \label{a9}
\end{equation}%
where the constant coupling $\lambda _{0}=-4E_{J}^{0}\cos \left( \frac{\pi
\Phi _{x}}{\Phi _{0}}\right) \left( \frac{\pi B\ell x_{0}}{\Phi _{0}}\right)
$ and the effective energy $\omega _{0}=8E_{c}\left( N_{1}-\frac{1}{2}%
\right) .$ In the rotating wave approximation the above Hamiltonian results
as
\begin{equation}
\hat{H}=\omega \hat{a}^{\dagger }\hat{a}+\frac{1}{2}\omega _{0}\hat{\sigma}%
_{z}+\lambda _{0}(\hat{\sigma}_{+}\hat{a}+\hat{a}^{\dagger }\hat{\sigma}%
_{-}).
\end{equation}

Now, in the interaction picture the Hamiltonian is written as, $\hat{H}_{I}=%
\hat{U}_{0}^{\dagger }\hat{H}\hat{U}_{0}-i\hbar \hat{U}_{0}^{\dagger }\frac{%
\partial \hat{U}_{0}}{\partial t},$ where $\hat{U}_{0}=\exp \left[ -i\left(
\omega \hat{a}^{\dagger }\hat{a}+\frac{\omega _{0}\hat{\sigma}_{z}}{2}%
\right) t\right] $ is the evoluion operator. Assuming the system operating
under the resonant condition, i.e., $\omega =\omega _{0}$, and setting $\hat{%
\sigma}_{z}=\hat{\sigma}_{+}\hat{\sigma}_{-}-\hat{\sigma}_{-}\hat{\sigma}%
_{+} $ and $\hat{\sigma}_{\pm }=$ $\left( \hat{\sigma}_{x}\pm i\hat{\sigma}%
_{y}\right) /2\ ,$ with \ $\hat{\sigma}_{y}=(\left\vert e\right\rangle
\left\langle g\right\vert -\left\vert e\right\rangle \left\langle
g\right\vert )/i$ the interaction Hamiltonian is led to the abbreviated form,

\begin{equation}
\hat{H}_{I}=\beta \left( \hat{a}^{\dagger }\hat{\sigma}_{-}+\hat{a}\hat{%
\sigma}_{+}\right) ,  \label{a11}
\end{equation}%
where $\beta =-\lambda _{0},$ $\hat{\sigma}_{+}$ $(\hat{\sigma}_{-})$ is the
raising (lowering) operator for the CPB.

We note that the coupling constant $\beta $ can be controlled through the
flux $\Phi _{x}$, which influences the mentioned small loops\emph{\ }in the
left and right places. Furthermore, we can control the gate charge $N_{1}$
via the gate voltage $V_{1}$ syntonized to the coupling. It should be
mentioned that the energy $\omega _{0}$ depends on the induced flux $\Phi
_{x}$. So, when we syntonize the induced flux $\Phi _{x}$ the energy $\omega
_{0}$ changes. To avoid unnecessary transitions during these changes, we
assume the changes in the flux being slow enough to obey the adiabatic
condition.

Next we show how to make holes in the statistical distribution of
excitations in the NR. We start from the CPB initially prepared in its
ground state $\left\vert CPB\right\rangle =\left\vert g\right\rangle ,$ and
the NR initially prepared in the coherent state, $\left\vert NR\right\rangle
=\left\vert \text{$\alpha $}\right\rangle .$Then the state $\left\vert \Psi
\right\rangle $ that describes the intire system (CPB plus NR) evolves as
follows
\begin{equation}
\left\vert \Psi _{NC}(t)\right\rangle =\hat{U}(t)\left\vert g\right\rangle
\left\vert \alpha \right\rangle ,  \label{h}
\end{equation}%
where $\hat{U}(t)=\exp (-it\hat{H}_{I})$ is the (unitary) evolution operator
and $\hat{H}_{I}$ is the interaction Hamiltonian, given in Eq. (\ref{a11}).

Setting $\hat{\sigma}_{+}=\left\vert g\right\rangle \left\langle
e\right\vert $ \ and $\hat{\sigma}_{-}=\left\vert e\right\rangle
\left\langle g\right\vert $\ we obtain after some algebra,

\begin{eqnarray}
\hat{U}(t) &=&\cos (\beta t\sqrt{\hat{a}^{\dagger }\hat{a}+1})\left\vert
g\right\rangle \left\langle g\right\vert \text{ }+\text{ }\cos (\beta t\sqrt{%
\hat{a}^{\dagger }\hat{a}})\left\vert e\right\rangle \left\langle
e\right\vert  \notag \\
&&-i\frac{\sin (\beta t\sqrt{\hat{a}^{\dagger }\hat{a}+1})}{\sqrt{\hat{a}%
^{\dagger }\hat{a}+1}}\hat{a}\left\vert g\right\rangle \left\langle
e\right\vert \text{ }-i\frac{\sin (\beta t\sqrt{\hat{a}^{\dagger }\hat{a}})}{%
\sqrt{\hat{a}^{\dagger }\hat{a}}}\hat{a}^{\dagger }\left\vert e\right\rangle
\left\langle g\right\vert .
\end{eqnarray}%
In this way, the evolved state in Eq.(\ref{h}) becomes%
\begin{equation}
\left\vert \Psi _{NC}(t)\right\rangle =e^{-\frac{\left\vert \alpha
\right\vert ^{2}}{2}}\sum_{n=0}^{\infty }\frac{\alpha ^{n}}{\sqrt{n!}}[\cos
(\omega _{n}\tau )\left\vert g,n\right\rangle \text{ }-i\sin (\omega
_{n}\tau )\left\vert e,n+1\right\rangle ],  \label{h1}
\end{equation}%
where $\omega _{n}=\beta \sqrt{n+1}$, If we detect the CPB in the state $%
\left\vert g\right\rangle $ after a convenient time interval $\tau _{1}$
then the state $\left\vert \Psi _{NC}(t)\right\rangle $ reads
\begin{equation}
\left\vert \Psi _{NC}(\tau _{1})\right\rangle =\eta _{1}\sum_{n=0}^{\infty }%
\frac{\alpha ^{n}}{\sqrt{n!}}\cos (\omega _{n}\tau _{1})\left\vert
n\right\rangle ,  \label{h2}
\end{equation}%
where $\eta _{1}$ is a normalization factor. If we choose $\tau _{1}$ in a
way that $\beta \sqrt{n_{1}+1}\tau _{1}=\pi /2$, the component $\left\vert
n_{1}\right\rangle $ in the Eq.(\ref{h2}) is eliminated.

In a second step, supose that this first CPB is rapidly substituted by
another one, also in the initial state $\left\vert g\right\rangle$, that
interacts with the NR after the above detection. For the second CPB the
initial state of the NR is the state given in Eq.(\ref{h2}), produced by the
detection of the first CPB in $\left\vert g\right\rangle $. As result, the
new CPB-NR system  evolves to the state
\begin{equation}
\left\vert \Psi _{NC}(\tau _{2})\right\rangle =\sum_{n=0}^{\infty }\frac{%
\alpha ^{n}}{\sqrt{n!}}[\cos (\omega _{n}\tau _{2})\cos (\omega _{n}\tau
_{1})\left\vert g,n\right\rangle -i\cos (\omega _{n}\tau _{1})\sin (\omega
_{n}\tau _{2})\left\vert e,n+1\right\rangle ].  \label{h3}
\end{equation}%
\qquad \qquad

Next, the detection of the second CPB again in the state $\left\vert
g\right\rangle $ leads the entire system collapsing to the state
\begin{equation}
\left\vert \Psi _{NC}(\tau _{2})\right\rangle =\eta _{2}\sum_{n=0}^{\infty }%
\frac{\alpha ^{n}}{\sqrt{n!}}[\cos (\omega _{n}\tau _{2})\cos (\omega
_{n}\tau _{1})\left\vert n\right\rangle ],  \label{h4}
\end{equation}%
where $\eta _{2}$ is a normalization factor. In this way, the choice $\beta
\sqrt{n_{2}+1}\tau _{2}=\pi /2$ makes a second hole, now in the component $%
\left\vert n_{2}\right\rangle $.

\begin{figure}[!h]
\subfigure[][]{
\includegraphics[{height=5.cm,width=5.0cm}]{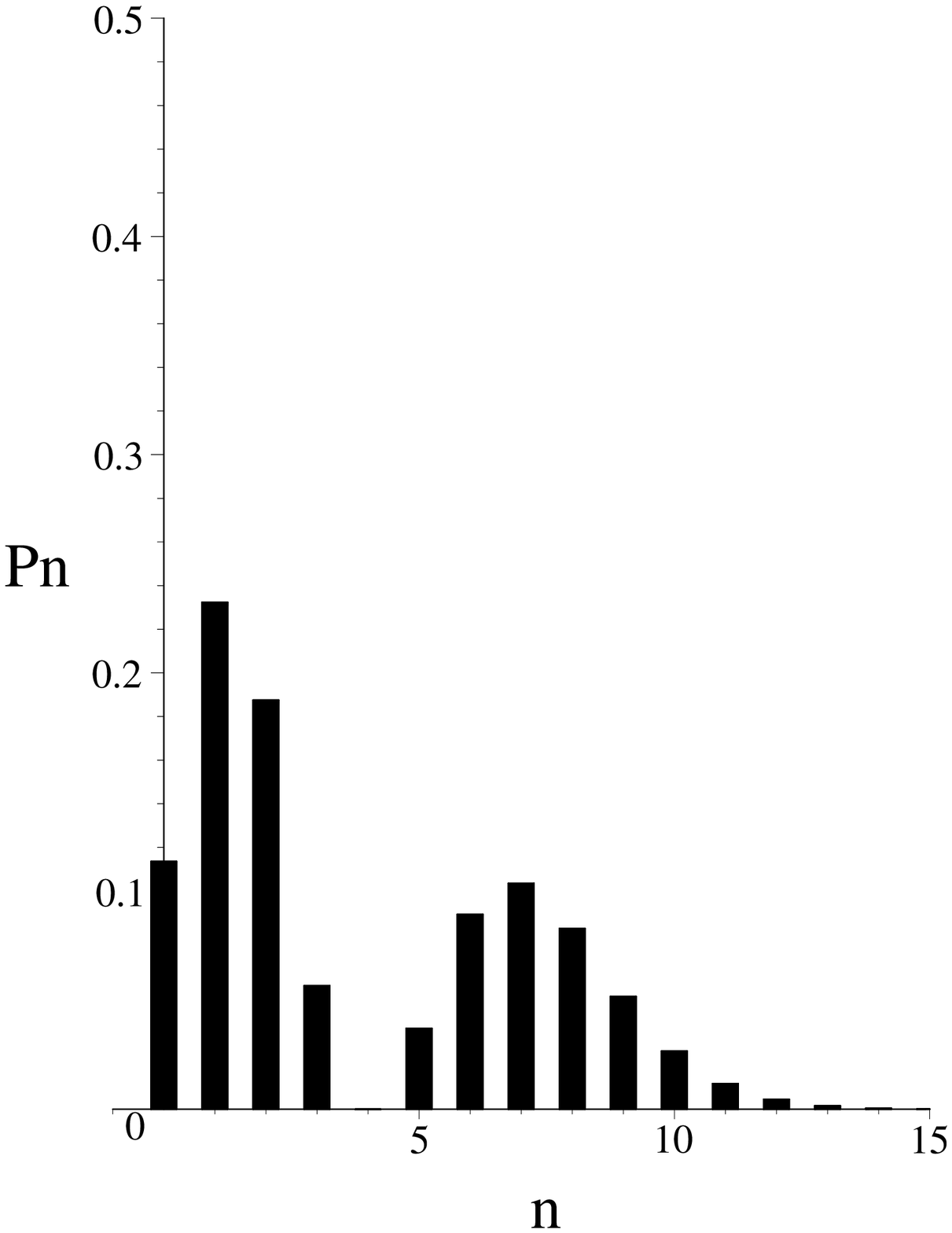}
} \subfigure[][]{
\includegraphics[{height=5cm,width=5.0cm}]{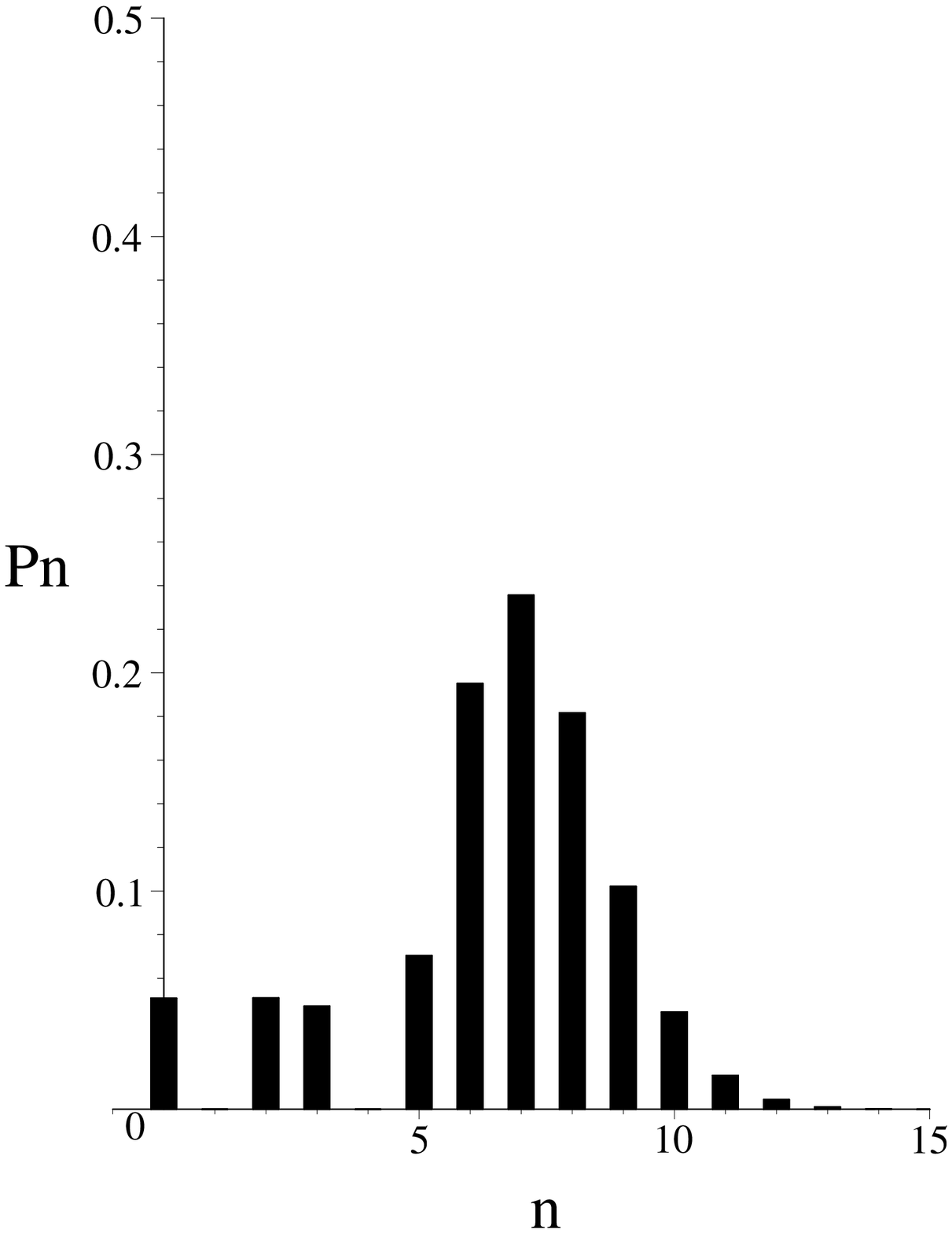}
} \subfigure[][]{
\includegraphics[{height=5cm,width=5.0cm}]{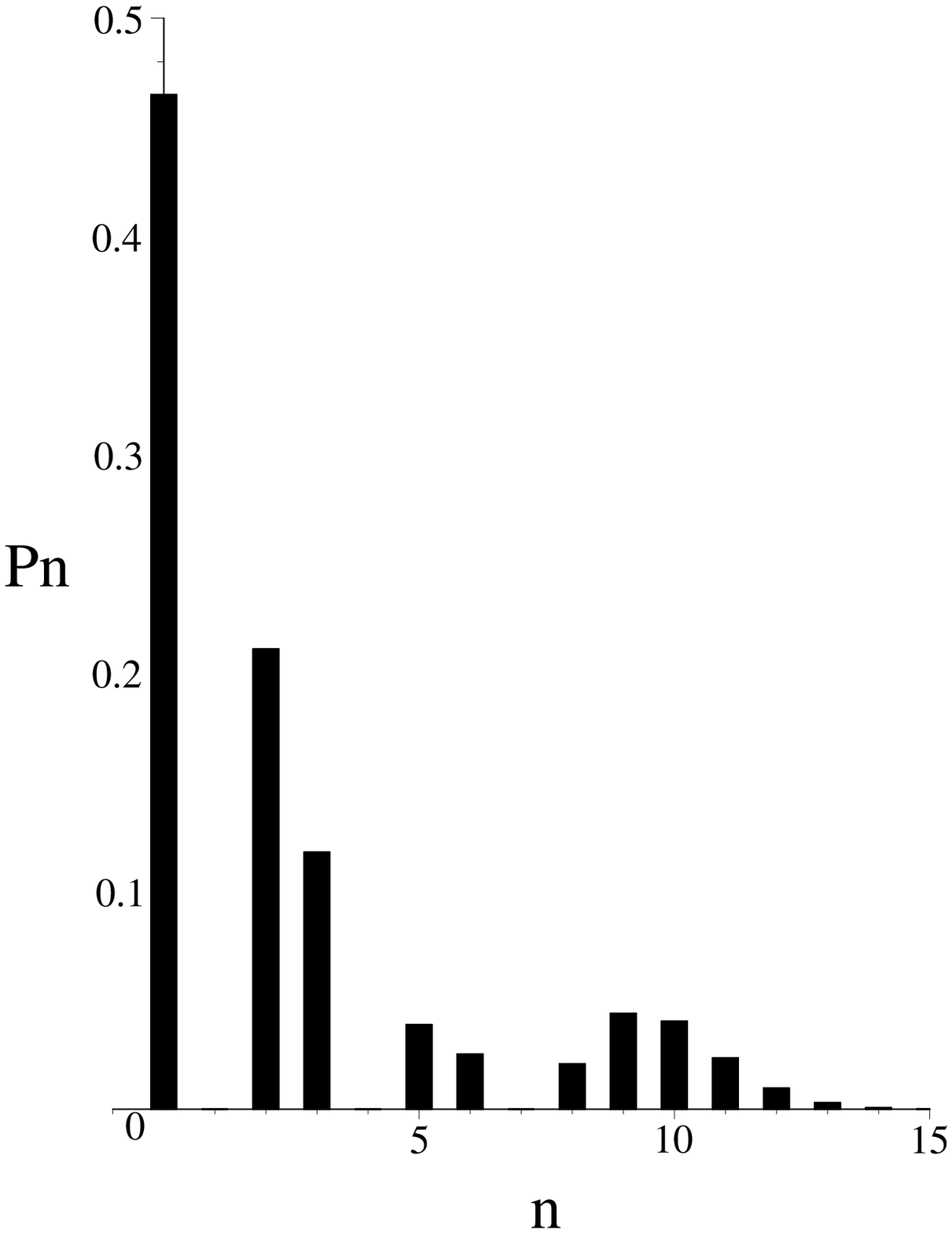}
}
\caption{Holes in the photon number distribution, for $\protect\alpha=2.0$,
(a) at $n_1=4$, for the $1^{st}$ step; (b) at $n_1=4$ and $n_2=1$, for the $%
2^{nd}$ step; (c) at $n_1=4$, $n_2=1$ and $n_3=7$, for the $3^{rd}$ step.}
\label{figure1}
\end{figure}

By repeating this procedure $M$ times we obtain the generalized result for
the $M-th$ CPB detection as%
\begin{equation}
\left\vert \Psi _{NC}(\tau _{M})\right\rangle =\eta _{M}\sum_{n=0}^{\infty }%
\frac{\alpha ^{n}}{\sqrt{n!}}\prod\limits_{j=1}^{M}\cos (\omega _{n}\tau
_{j})\left\vert n\right\rangle ,  \label{h5}
\end{equation}%
where $\tau _{j}$ is the $j-th$\ CPB-NR interaction time. According to the
Eq (\ref{h5}) the number of CPB being detected coincides with the number of
holes produced in the statistical distribution. In fact the Eq (\ref{h5})
allows one to find the expression for the statistical distribution, $%
P_{n}=\left\vert \langle n|\Psi _{NC}(\tau _{M})\right\vert ^{2};$ a little
algebra furnishes
\begin{equation}
P_{n}=\frac{(\alpha ^{2n}/n!)\prod_{j=1}^{M}cos^{2}(\omega _{n}\tau _{j})}{%
\sum_{m=0}^{\infty }(\alpha ^{2m}/m!)\prod_{j=1}^{M}cos^{2}(\omega _{n}\tau
_{j})},  \label{buracos}
\end{equation}%
To illustrate results we have plotted the Fig.(\ref{figure1}) showing the
controlled production of holes in the photon number distribution.

The success probability to produce the desired state is given by
\begin{equation}
P_{s}=e^{-\left\vert \alpha \right\vert ^{2}}\sum_{m=0}^{\infty }(\alpha
^{2m}/m!)\prod_{j=1}^{M}cos^{2}(\omega _{n}\tau _{j}).  \label{ps}
\end{equation}%
Note that the holes exhibited in Fig.(\ref{figure1})(a), \ref{figure1}(b),
and \ref{figure1}(c) occur with success probability of $9\%$, $4\%$, and $%
0.3\%$, respectively.

We can take advantage of the this procedure applying it to the engineering
of nonclassical states, e.g., to prepare Fock states \cite{Escher05PRA} and
their superpositions \cite{Aragao04PLA}. To this end, we present two
strategies: in the first we eliminates the components on the left and right
sides of a desired Fock state $|N\rangle $, namely: $|N-1\rangle ,$ $%
|N-2\rangle ,...$and $|N+1\rangle ,|N+2\rangle ,...;$ in the second one, we
only eliminate the left side components of a desired Fock state $|N\rangle $%
. In both cases, it is convenient to consider the final state of the NR as,
\begin{equation}
\left\vert \Psi _{NC}(\tau _{M})\right\rangle ^{\prime }=\eta _{M}^{\prime
}\sum_{n=0}^{\infty }\frac{\alpha ^{n}}{\sqrt{n!}}(-i)^{M}\prod%
\limits_{j=1}^{M}\sin (\omega _{n+j}\tau _{j})\left\vert n+M\right\rangle ,
\label{h6}
\end{equation}%
which is easily obtained by detecting the Cooper pair box in the state $%
|e\rangle $. The success probability $P_{s}^{\prime }$ to produce a Fock
state $|N\rangle $ reads
\begin{equation}
P_{s}^{\prime }=e^{-\left\vert \alpha \right\vert ^{2}}\sum_{m=0}^{\infty
}(\alpha ^{2m}/m!)\prod_{j=1}^{M}sin^{2}(\omega _{n+j}\tau _{j}).
\label{pss}
\end{equation}

In the first strategy, we prepare Fock states $|N\rangle $ with $N=M$, i.e.,
the phonon-number $N$ coincides with the number of CPB detections $M$. The
fidelity of these states is given by the phonon number distribution at $%
P_{M} $ \ associated with the state $\left\vert \Psi _{NC}(\tau
_{M})\right\rangle ^{\prime },$%
\begin{equation}
P_{M}=\frac{\prod_{j=1}^{M}\sin ^{2}(\sqrt{j}\beta \tau _{j})}{%
\sum_{n=0}^{\infty }(\alpha ^{2n}/n!)\prod_{n=1}^{M}\sin ^{2}(\sqrt{n+j}%
\beta \tau _{n})}.  \label{ps1}
\end{equation}

\begin{figure}[!h]
\subfigure[N4][]{
\includegraphics[{height=5.cm,width=5.0cm}]{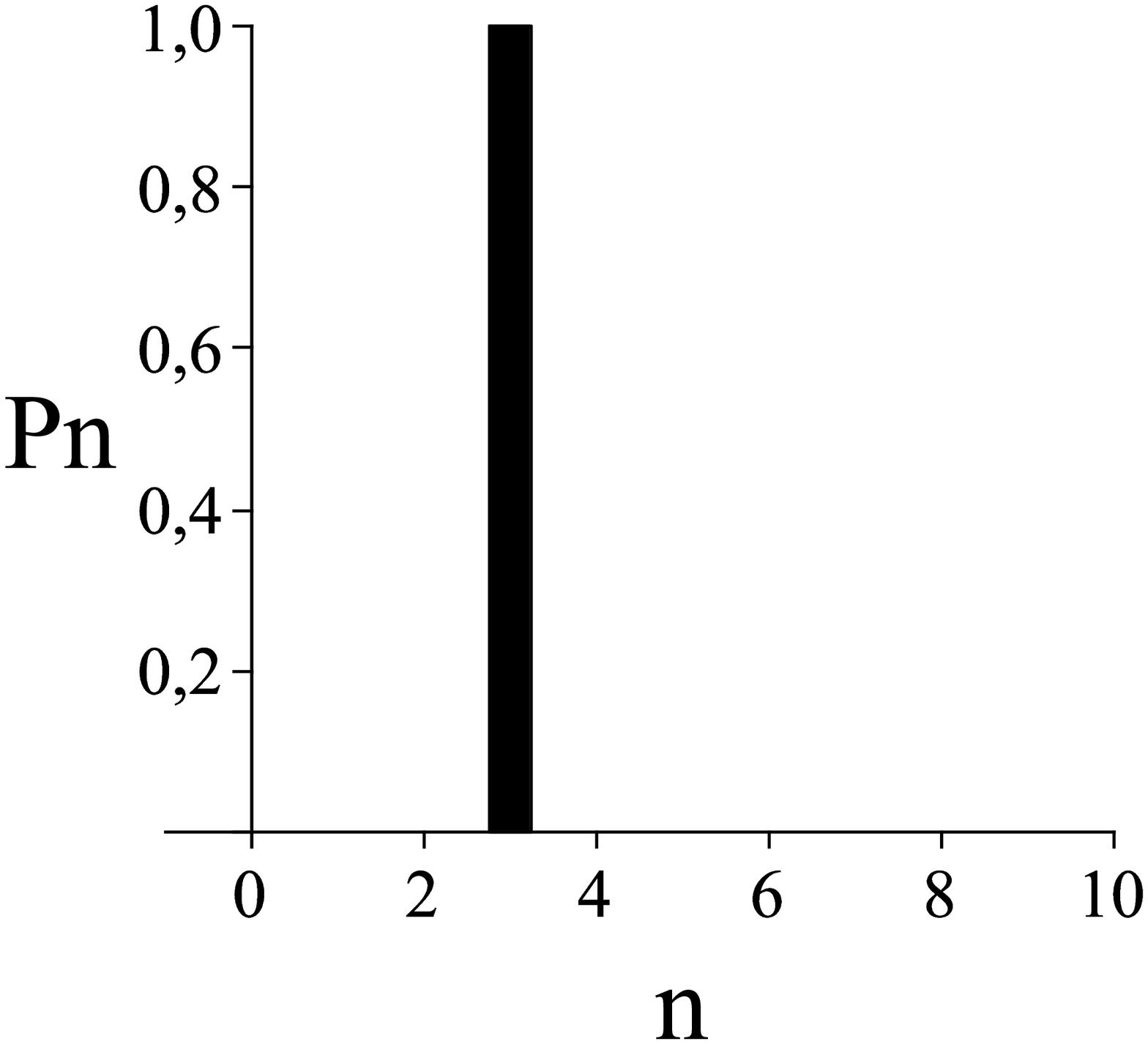}
} \subfigure[N5][]{
\includegraphics[{height=5cm,width=5.0cm}]{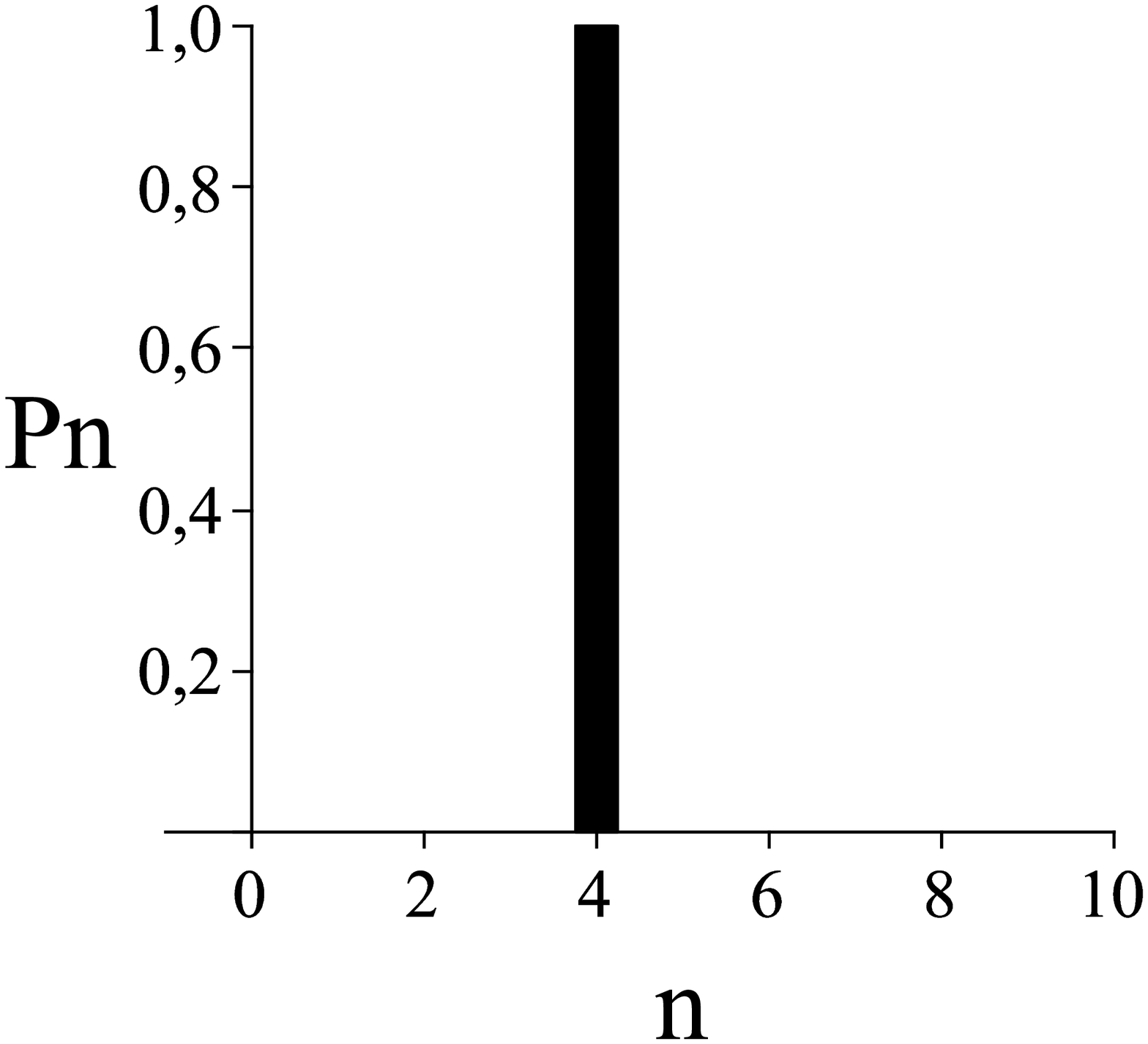}
} \subfigure[N6][]{
\includegraphics[{height=5cm,width=5.0cm}]{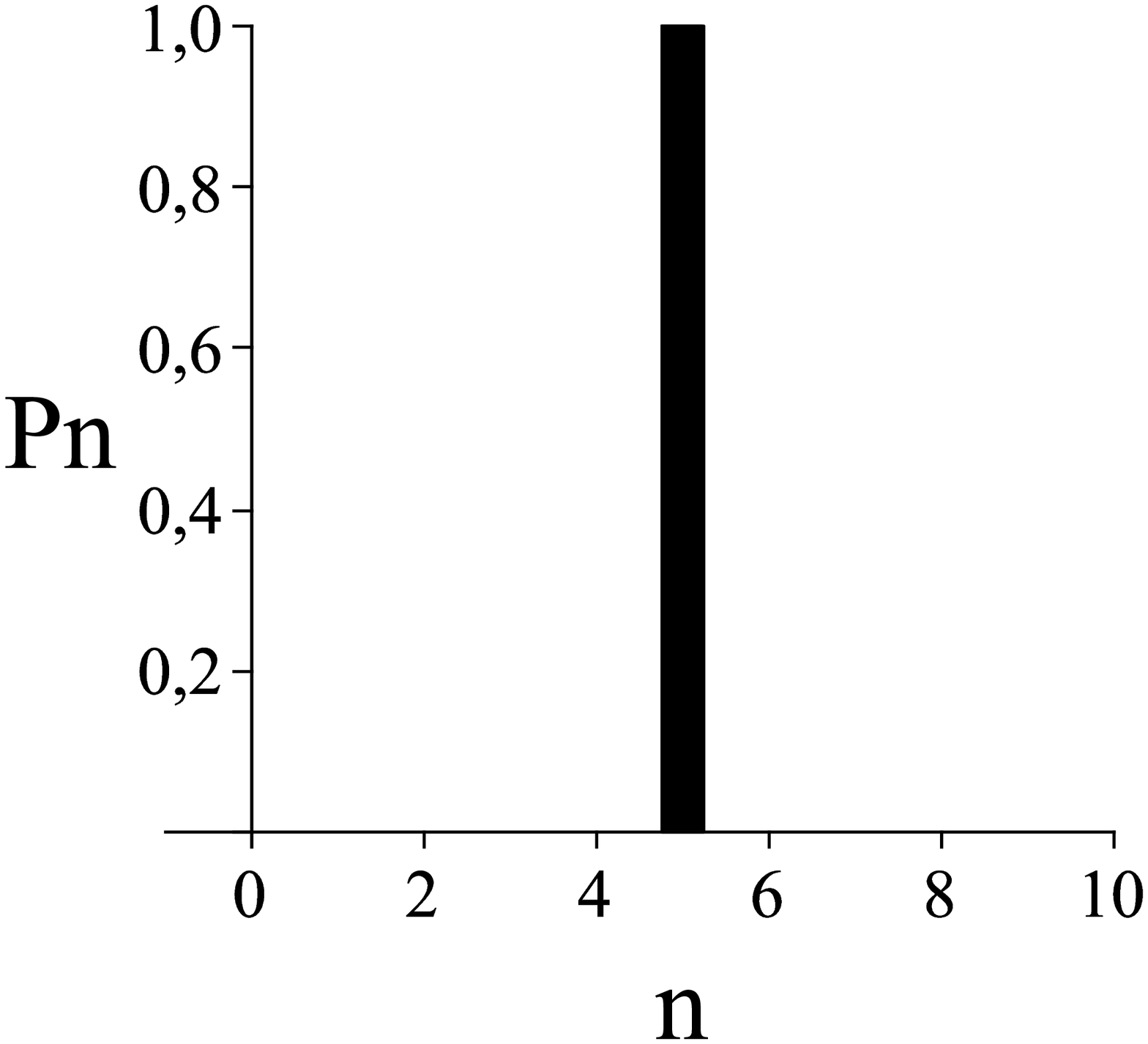}
}
\caption{Phonon number distribution exhibiting the creation of Fock state:
(a) $|3\rangle$ ($P_{s}^\prime=17\%$), (b) $|4\rangle$ ($P_{s}^\prime=11\%$%
), and (c) $|5\rangle$ ($P_{s}^\prime=7\%$); all with fidelity of $99\% $
and initial coherent state with $\protect\alpha=0.6$.}
\label{Number}
\end{figure}

We note that, in this case the fidelity coincides with the $N-th$ component
of the statistical distribution $Pn$. The Fig.(\ref{Number}) shows the
phonon-number distribution exhibiting the creation of Fock state $|3\rangle $%
, $|4\rangle $, and $|5\rangle $; all with fidelity of $99\%$, for an
initial coherent state with $\alpha =0.6$.

\begin{figure}[!h]
\subfigure[N4][]{
\includegraphics[{height=5.cm,width=5.0cm}]{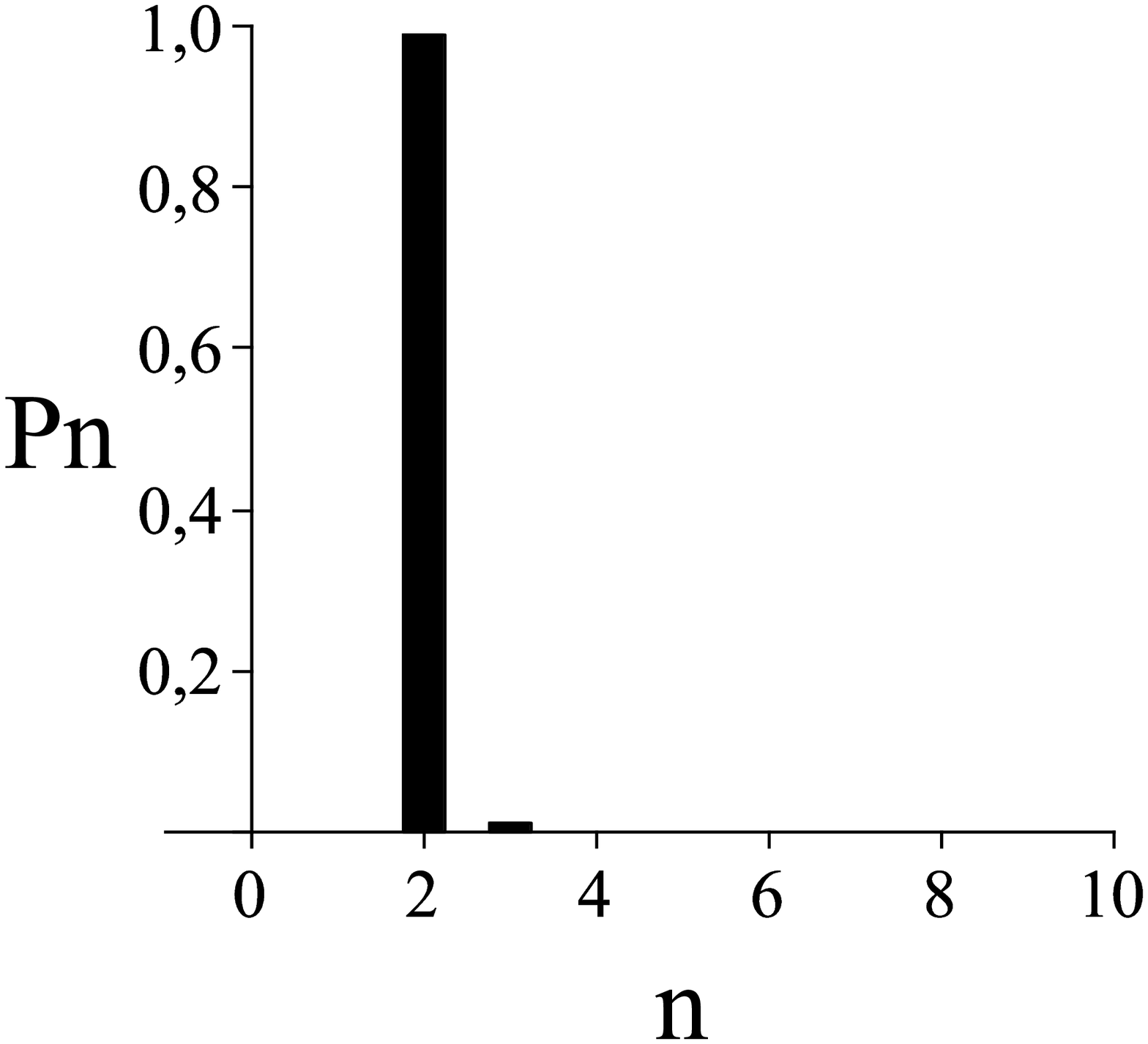}
}
\subfigure[N5][]{
\includegraphics[{height=5cm,width=5.0cm}]{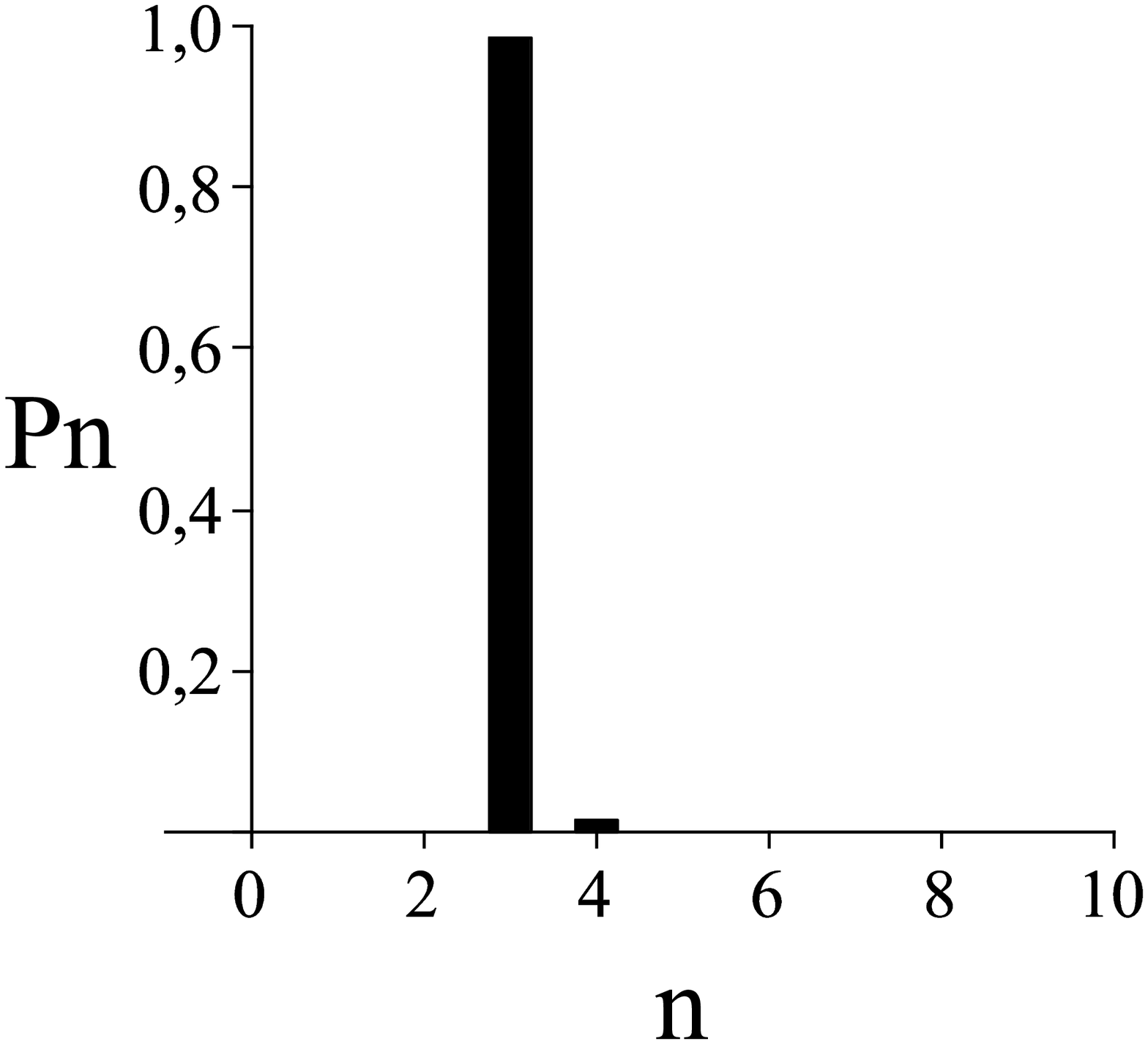}
}
\subfigure[N6][]{
\includegraphics[{height=5cm,width=5.0cm}]{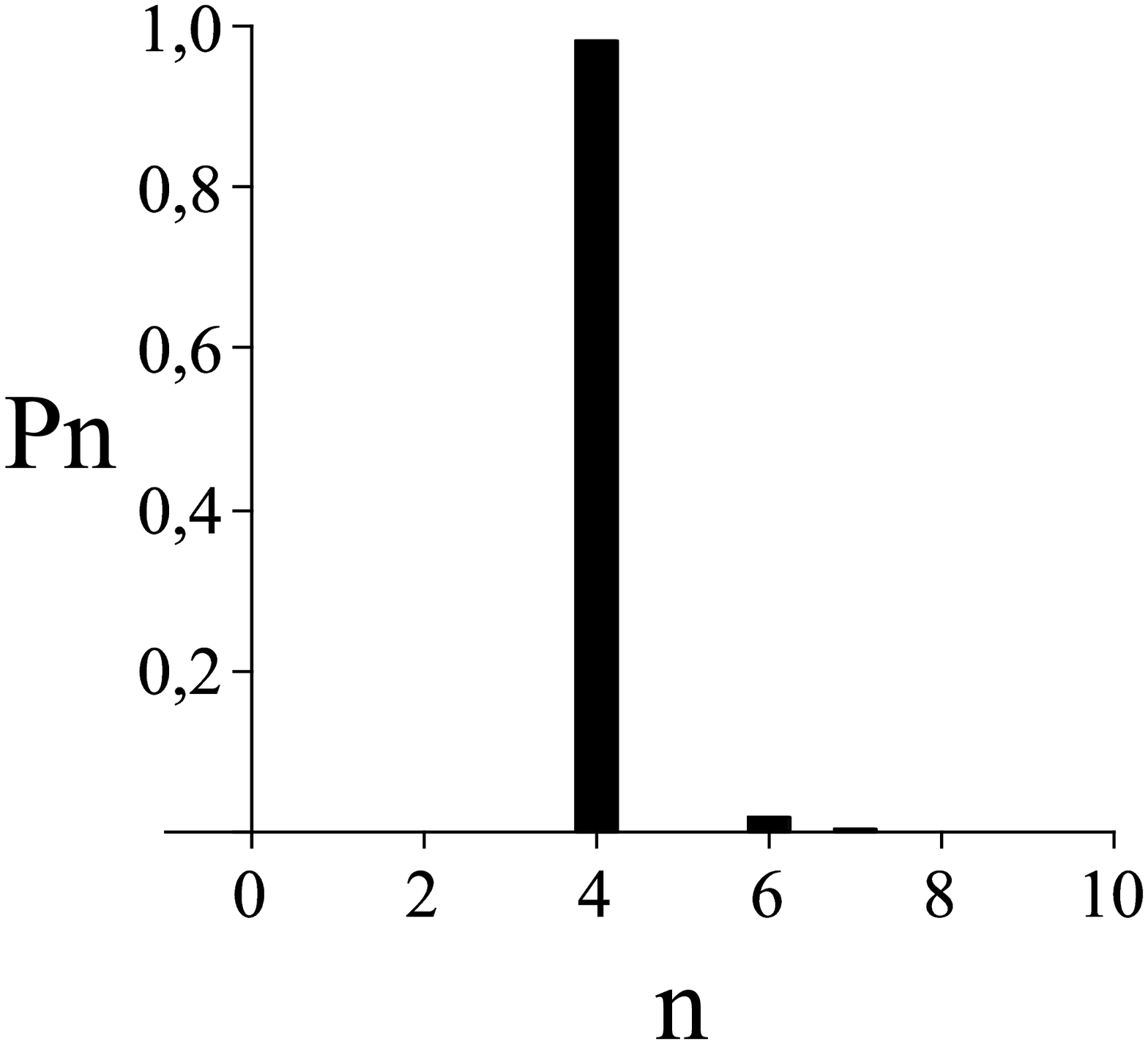}
}
\caption{Phonon number distribution exhibiting the creation of Fock state:
(a) $|2\rangle$ ($P_{s}^\prime=17\%$), (b) $|3\rangle$ ($P_{s}^\prime=1\%$),
and (c) $|4\rangle$ ($P_{s}^\prime=0.3\%$); all with fidelity of $98\% $ and
initial coherent state with $\protect\alpha=0.6$.}
\label{number2}
\end{figure}

In the second strategy, we prepare Fock states $|N\rangle $ with $N=2M$ or $%
2M-1$. The associated fidelity is also given by the Eq.(\ref{ps1}). The Fig.(%
\ref{number2}) shows the phonon-number distribution exhibiting the creation
of Fock states $|3\rangle $, $|4\rangle $, and $|5\rangle $, all them with
same fidelity $99\%,$ for an initial coherent state with $\alpha =0.6$.

\section{Conclusion}

Concerning with the feasibility of the scheme, it is worth mentioning some
experimental values of parameters and characteristics of our system: the
maximum value of the coupling constant $\beta _{m\acute{a}x}\approx 45MHz$,
with $B\approx 0,1T$, $\ell =30\mu m$, $x_{0}=500fm$ and $E_{J}^{0}=5GHz$,
with $\omega _{0}=200\pi MHz$.\emph{\ \cite%
{huang,a16a,a33,a34,b1,b2,bb2,tr,b3}}. The expression choosing the time
spent to make a hole, $\beta \sqrt{n_{j}+1}\tau _{j}=\pi /2,$ funishes $\tau
_{j}\simeq $ $0.3~ns,$ when assuming all the CPB\ previously prepared at $t=0
$. On the other hand, the decoherence times of the CPB and the NR are
respectively $500~ns$ and $160$ $\mu s$ \cite{tr}. Accordingly, one may
create about 1600 holes before the destructive action of decoherence.
However, when considering the success probability to detect all CPB in the
state $\left\vert g\right\rangle $,\ a more realistic estimation drastically
reduces the number of holes. A similar situation occurs in \cite%
{malboi,ard1,ard}\emph{,} using atom-field system to make holes in the
statistical distribution $P_{n}$ of a field state; in this case, about $1\mu
s$ is spent to create a hole whereas $1ms$\ is the decoherence time of a
field state inside the cavity. So, comparing both scenarios the present
system is about 60\% more efficient in comparison with that using the
atom-field system. Concernig with the generation of a Fock state $\left\vert
N\right\rangle $, it is convenient starting with a low excited initial
(coherent) state, which involves a low number of Fock components to be
deleted via our hole burning procedure. According to the Eq. (\ref{ps1})
when one must delete many components of the initial state to achieve the
state $\left\vert N\right\rangle $ this drastically reduces the success
probability. As consequence, this method will work only for small values of $%
N$ ( $N\lesssim 5$ ).

\section{Acknowledgements}

The authors thank the FAPEG (CV) and the CNPq (ATA, BB) for partially
supporting this work.

\end{document}